\documentclass[dvips]{article}

\usepackage{icrc2011}
   \usepackage{url}
\title{Target of opportunity observations of flaring blazars with VERITAS}

\newcommand{\etal}{\MakeLowercase{et al. }} 

\newcommand{\F}{{\em Fermi} }
\newcommand{\FL}{{\em Fermi}-LAT }
\shorttitle{M. Errando \etal VERITAS observations of flaring blazars}

\authors{M. Errando$^{1}$ for the VERITAS collaboration$^{2}$}
\afiliations{$^1$Department of Physics and Astronomy, Barnard College, Columbia University, NY 10027, USA\\
$^2$ see J. Holder et al. (these proceedings) or \url{http://veritas.sao.arizona.edu/conferences/authors?icrc2011}}
\email{errando@astro.columbia.edu}

\abstract{VERITAS is an array of four imaging atmospheric Cherenkov telescopes that observes $\gamma$-ray sources in the very high energy range (VHE, $E>100$\,GeV). A large fraction of the known $\gamma$-ray sources are of extragalactic origin and belong to the blazar class: active galactic nuclei with relativistic jets pointing along the line of sight. $\gamma$-ray emission from blazars is typically variable. Part of the VERITAS observation program is devoted to follow-up triggers on flaring blazars, initiated by observations at other wavelengths. We will present those target of opportunity observations based on optical and GeV ({\em Fermi}-LAT) alerts performed with VERITAS during the past 2 seasons, and discuss the results in a multiwavelength context.}
\keywords{ VERITAS, AGN, Blazar, Gamma-ray, TeV, VHE, \F}

\begin{document}
\maketitle

\section{Introduction}
There are currently 48 extragalactic objects detected at VHE energies \cite{tevcat}. 42 of those are blazars: active galactic nuclei (AGN) with a relativistic jet pointing along the line of sight to the observer. Blazars show broadband non-thermal emission extending from radio frequencies to $\gamma$ rays, with a typical two-bump spectral energy distribution (SED). If the emission is assumed to be originated by relativistic electrons, which is the most-accepted scenario, the low-energy component is explained as synchrotron radiation from the electrons in the jet magnetic field, while the high-energy component is attributed to inverse-Compton scattering of low energy photons off the relativistic electrons.  According to their optical properties, blazars are classified as Flat Spectrum Radio Quasars (FSRQs) when line emission dominates the optical spectrum, or BL Lac-type blazars when the optical emission is continuum dominated. The frequency where the synchrotron component of BL Lacs has its maximum is used to further classify them as Low, Intermediate and High-frequency peaked BL Lacs (LBL, IBL, and HBL) when the the peak is located respectively in the infrared, optical, or X-ray band. Most of the VHE detected blazars (30) fall into the HBL subclass.

Blazars are typically variable at all wavelengths, with the fastest and most dramatic variability appearing in the $\gamma$-ray band. About a third of all VHE blazars have been detected only during flaring states. Therefore, it is natural to use ground-based $\gamma$-ray telescopes to follow-up on flares from blazar candidates to try to observe them when they are in a high state and the chances of detection are larger. VERITAS currently has an existing program using public \FL data to identify high GeV $\gamma$-ray states \cite{monitor}. VERITAS also responds to public announcements of blazar flares in the $\gamma$ ray, X ray, and optical band, in addition to an internal blazar monitoring program at optical frequencies at the UCO/Lick  Observatory.

\section{Observations and results}
VERITAS is an array of four imaging atmospheric Cherenkov telescopes located at the Fred Lawrence Whipple Observatory (FLWO) in southern Arizona (31 40\,N, 110 57\,W,  1268\,m a.s.l.). It combines a large effective area over a large range of energies (100\,GeV to 30\,TeV) with an energy resolution of 15-25\% and an angular resolution of less that $0.1^{\circ}$. The high sensitivity of VERITAS allows the detection of sources with a flux of 0.01 times that of the Crab Nebula in about 25 hours. 

\begin{table*}[t]
\begin{footnotesize}
\begin{center}
\begin{tabular}{l|ccccccc}
\hline
& Type & z & Observations & Exposure & Trigger & $E_{th}$ & $F (E>E_{th})$\\
& & & & & & GeV &$10^{-12}$cm$^{-2}$s$^{-1}$ \\
\hline
QSO 0133+476 & FSRQ & 0.859 & 02/2008 & 0.8\,h & optical & 360 & $< 12 $\\
B2 0200+30 & AGN & & 01/2011 & 1.5\,h & \F & 200 & $< 7.2 $\\
CGRaBS J0211+1051 & BL Lac & 0.200 &01-02/2011 & 4.0\,h & \F & 200 & $< 9.7 $\\
AO 0235+16 & LBL & 0.940 &  09-10 / 2008 & 4.3\,h  & \F & 170 & $< 2.6 $\\
4C +55.17 & FSRQ & 0.896 & 5,11-12/2010, 01/2011& 17.7\,h & \F $\star$& 200 & $< 2.6 $\\
SBS 1150+497 & FSRQ & 0.330 & 5/2011& 4.2\,h &\F & 220 & $< 6.1  $\\
4C 21.35 & FSRQ & 0.435 & 12/2009, 05/2010, 03/2011& 23.3\,h & {\em Fermi}, optical& 180 & $< 4.0 $\\
3C 279 & FSRQ & 0.536 &05-06/2011&  3.5\,h &\F & 280 & $< 5.5  $\\
PKS 1510-089 & FSRQ & 0.360 & 03-04/2009 & 4.3\,h &  {\em AGILE}, \F & 360 & $< 2.5 $\\
GB6 J1700+6830 & FSRQ & 0.301 & 03/2009 & 0.8\,h & \F & 300 & $< 9.1  $\\
3C 454.3 & FSRQ & 0.859 & 09/2009, 11-12/2010 & 1.8\,h & \F & 200 & $< 3.2 $\\
\hline
\end{tabular}
\caption{List of ToO VERITAS observations of blazars. The object type, redshift, month and year of observations, total exposure and trigger source are indicated. \\
{\small ($\star$) this source was observed after indications of a promising hard spectrum source in the \F data, but not particularly because the source was in a high flux state.}}\label{table}
\end{center}
\end{footnotesize}
\end{table*}

A summary of VERITAS observations of flaring blazars can be found in Table~\ref{table}, together with the flux upper limits resulting from the VERITAS exposure. This report does not include observations of known VHE BL Lac-type blazars during flares, which will be reported elsewhere. Some remarks on selected sources follow.

\subsection{QSO 0133+476}
QSO 0133+476 is a FSRQ at redshift $z=0.859$ that was briefly observed in February 2008 with VERITAS after a bright optical flare was reported \cite{yoshida}. The observations lasted 0.8\,h and resulted on a flux upper limit of $F(E>360\,\mathrm{GeV}) < 12 \times 10^{-12}\,\mathrm{cm}^{-2}\,\mathrm{s}^{-1}$.

\subsection{B2 0200+30}
B2 0200+30 is an AGN of unidentified type associated with the \FL source 1FGL~J0203.5+3044. This source was identified to be in a high state in the GeV band using a VERITAS automated tool for rapid all-sky analysis of \FL data \cite{monitor} in December 2010. The $\gamma$-ray source was found to have a flux $\sim 40$ times higher than its average spectrum during the first year of the \F mission \cite{1fgl}. A GeV spectrum
was derived using \F data during the flare and revealed a spectral hardening of the source. 

VERITAS observations were triggered after approval of a director's discretionary time request. Only
90 minutes of good quality data could be accumulated due to adverse weather conditions.

\subsection{CGRaBS J0211+1051}
CGRaBS J0211+1051 is a BL Lac-type blazar with a recently-determined redshift of $z=0.20 \pm 0.05$ \cite{cgrabs-z}. It was reported by the \FL team to be in a flaring state with a peak flux of about 25 times its flux in the 1FGL catalog in January 2011 \cite{cgrabs-fermi}. 4.0\,h of VERITAS observations were taken during the $\gamma$-ray flare, resulting in a flux upper limit of $F(E>200\,\mathrm{GeV}) < 9.7 \times 10^{-12}\,\mathrm{cm}^{-2}\,\mathrm{s}^{-1}$.

\subsection{AO 0235+16}
AO 0235+16 is a well-known LBL at a redshift of $z=0.940$ \cite{ao-z}. It was originally detected in $\gamma$ rays by EGRET \cite{3eg}. \FL reported a bright flare in September 2008 \cite{ao-fermi} followed by strong activity at short timescales \cite{ao-fermi-fast}.

VERITAS observed AO 0235+16 during the {\em Fermi}-LAT-reported high state for 4.3\,h, deriving a 99\% confidence level (c.l.) integral flux upper limit for $E>170$\,GeV of $2.6 \times 10^{-12}\,\mathrm{cm}^{-2}\,\mathrm{s}^{-1}$ \cite{veritas-ul}.

\subsection{4C +55.17}
4C +55.17 is a bright {\em Fermi}-LAT-detected FSRQ at $z=0.896$, associated with 1FGL~J0957.7+5523 \cite{1fgl}. However, its spectral index of $\Gamma = -2.05 \pm 0.03$ is much harder than that of the average LAT-detected FSRQs \cite{fermilac}. Quasars are also very variable objects in the $\gamma$-ray band \cite{variability}, but 4C +55.17 has not shown any significant flux variability throughout the \F mission. 

The very hard spectral index observed by \F makes this object an interesting target for ground-based $\gamma$-ray telescopes \cite{mcconville}. VERITAS has observed 4C +55.17 for 17.7 hours during 2010 and 2011, resulting on a flux upper limit of $F(E>200\,\mathrm{GeV}) < 2.6 \times 10^{-12}\,\mathrm{cm}^{-2}\,\mathrm{s}^{-1}$.

\subsection{SBS 1150+497}
SBS 1150+497 is a flat spectrum radio quasar at redshift $z=0.330$ that was seen at $\gamma$-ray state an order of magnitude brighter than average by \F in April 2011 \cite{hays} followed by a $\gamma$-ray re-brightening and an X-ray flare \cite{reyes}.

VERITAS could observe the source for 4.2 hours of effective time that resulted in a flux upper limit of $F(E>220\,\mathrm{GeV}) < 6.1 \times 10^{-12}\,\mathrm{cm}^{-2}\,\mathrm{s}^{-1}$.

\subsection{4C 21.35}
4C~21.35 (also known as PKS\,1222+216, $z=0.432$) is a $\gamma$-ray-emitting FSRQ. 

It is the most-likely counterpart to the EGRET-detected source 3G~J1224+2118 \cite{3eg}, and subsequently subjected to several $\gamma$-ray studies and multiwavelength modeling. During the first three months of the \F mission, 4C~21.35 was not bright enough to be included in the \FL Bright Source List \cite{latbsl}. It was detected, however, in the First LAT Catalog \cite{1fgl}, where it was associated with 1FGL~J1224.7+2121. A $\gamma$-ray flare from this source was noted in April 2009 \cite{longo}. It was followed in 2009 December by an even larger flare, seen by both the {\it AGILE} Gamma-ray Imaging Detector \cite{4c-agile} and the {\it Fermi}-LAT \cite{ciprini}. On April 24 2010, {\it Fermi} LAT detected a particularly strong GeV outburst from the object \cite{donato}. The \FL analysis performed by \cite{neronov} indicated that during the flare the $\gamma$-ray emission of 4C~21.35 extended up to observed photon energies greater than $100$\,GeV, i.e., up to the very-high-energy (VHE) band. 
A second huge GeV outburst was recorded by \F and {\it AGILE} in  June 2010 \cite{longo2010,agile2}. At the same time (17 June 2010), a prominent excess at $0.07-0.4$\,TeV photon energies was detected by the MAGIC telescope from the position of 4C~21.35 \cite{mose,magic}, establishing this source as the third FSRQ firmly detected in the VHE range by ground-based instruments besides 3C~279 and PKS~1510$-$089. 

VERITAS observed 4C~21.35 during all major $\gamma$-ray flaring episodes except for June 2010, when moonlight and weather conditions prevented the observations. Additionally, a bright optical flare in March 2011 triggered additional VERITAS observations, that add up to a total of 23.3\,h. The derived upper limit including all VERITAS observations is $F(E>180\,\mathrm{GeV}) < 4.0 \times 10^{-12}\,\mathrm{cm}^{-2}\,\mathrm{s}^{-1}$.

\subsection{3C 279}
The FSRQ 3C~279, located at a redshift of $z=0.536$, is one of the brightest extragalactic objects in the $\gamma$-ray sky \cite{kniffen93} 
and it is an exceptionally variable
source at various energy bands, including the $\gamma$-ray regime, where strong flares detected by EGRET in 1991 and 1996 \cite{Wehrle98}. Its detection at VHE $\gamma$-rays by the MAGIC telescope \cite{icrc,science} is the first detection of an FSRQ in this energy band as well as the farthest object detected at these energy band.

In May 2011 3C 279 was observed at a high flux state by \F, reaching a flux level close to the highest seen since the satellite was launched. VERITAS observed the source during the high state for 3.5 hours, reporting a flux upper limit of $F(E>280\,\mathrm{GeV}) < 5.5 \times 10^{-12}\,\mathrm{cm}^{-2}\,\mathrm{s}^{-1}$.

\subsection{PKS~1510-089}
PKS~1510-089 was already detected in $\gamma$-rays by EGRET \cite{3eg} and 
exhibited very interesting activity at all wavelengths. It was also detected by 
AGILE during ten days of pointed observations from 2007-08-23 to 2007-09-01 \cite{Pucella}. 
In the period 2008-09, PKS~1510-089 was observed to be bright 
and highly variable in several frequency bands. A bright $\gamma$-ray state from PKS~1510-089 was observed in March and April 2009 by both \textit{Fermi}-LAT and AGILE 
\cite{1510-2,1510-3}. High states in X-rays and 
in optical were also reported during that period. In a recent paper, Marscher \etal \cite{marscher-1510} presented data from a multi-wavelength
campaign concerning the same flaring epoch. In that paper, the authors focus on analysis of the parsec-scale 
behavior and correlation of rotation of the optical polarization angle 
with the dramatic $\gamma$-ray activity.

VERITAS observed PKS~1510-089 during the flaring episodes of March and April 2009 following the {\em AGILE} and \F alerts. A total of 4.3\,h of data were taken where no significant signal was found. A 99\% c.l. integral flux upper limit for $E>360$\,GeV of $2.5 \times 10^{-12}\,\mathrm{cm}^{-2}\,\mathrm{s}^{-1}$ was derived from these observations \cite{veritas-ul}.

\subsection{GB6 J1700+6830}
GB6 J1700+6830 was first detected in the $\gamma$-ray band by \FL during a flare in March 2009 \cite{fermi-gb6}. The source was later associated to 1FGL J1700.1+6830 \cite{1fgl,fermilac}. This object is a known FSRQ at a redshift of $z=0.301$. VERITAS observed GB6 J1700+6830 following the \F detection for 0.8\,h, placing a 99\% c.l. upper limit
to the integral flux in $\gamma$ rays at $E>300$\,GeV of $9.1 \times 10^{-12}\,\mathrm{cm}^{-2}\,\mathrm{s}^{-1}$ \cite{veritas-ul}.

\subsection{3C 454.3}
3C 454.3 is a well-studied FSRQ detected several times in the $\gamma$-ray band by EGRET \cite{3eg}. Since the first days after the launch of {\em Fermi}, 3C 454.3 has been one of the brightest and most active sources in the $\gamma$-ray sky. VERITAS observed this object after a flare reported by \FL in September 2009 \cite{454-09} and during a major flare in late 2010 \cite{454-10} when it became the brightest object ever detected in the $\gamma$-ray band. During these two periods of observation 1.8\,h of good quality data were recorded. The resulting flux upper limit is $F(E>200\,\mathrm{GeV}) < 3.2 \times 10^{-12}\,\mathrm{cm}^{-2}\,\mathrm{s}^{-1}$.

\section{Conclusions}
The VERITAS blazar discovery program uses all available multiwavelength information to try to observe VHE blazar candidates when they are in high flux states. Target of opportunity observations will continue in the future, relying on public notifications (such as Astronomer's Telegrams), monitoring of available public light curves (optical, {\em Fermi}), and internal efforts on monitoring blazars in optical at the UCO/Lick Observatory and in $\gamma$ rays by automated analysis of the public \FL data.

\section{Acknowledgment}
This research is supported
by the NASA grant  NNX10AP66G. VERITAS is supported by grants from the US Department of Energy, the US National Science Foundation, and the
Smithsonian Institution, by NSERC in Canada, by Science Foundation Ireland, and by STFC in the UK. We acknowledge the
excellent work of the technical support staff at the FLWO and at the collaborating institutions in the construction and
operation of the instrument.


\clearpage

\end{document}